\documentclass[a4paper,12pt]{article}

\newcommand{\sect}[1]{\setcounter{equation}{0}\section{#1}}

\begin{document}
\topmargin 0pt
\oddsidemargin 0mm

\renewcommand{\thefootnote}{\fnsymbol{footnote}}
\newcommand{\g}{\bf G_n}
\begin{titlepage}
\begin{flushright}
OU-HET 385\\
INJE-TP-01-05 \\
hep-th/0105070
\end{flushright}

\vspace{5mm}
\begin{center}
{\Large \bf Bekenstein Bound, Holography and
Brane Cosmology in Charged Black Hole Backgrounds}
\vspace{12mm}

{\large
Rong-Gen Cai\footnote{email address: cai@het.phys.sci.osaka-u.ac.jp}
, Yun Soo Myung\footnote{email address: ysmyung@physics.inje.ac.kr}
and  Nobuyoshi Ohta\footnote{email address: ohta@phys.sci.osaka-u.ac.jp}} \\
\vspace{8mm}
{\em $^\ast$ $^\ddagger $ Department of Physics, Osaka University,
Toyonaka, Osaka 560-0043, Japan \\
 $^\dagger $ Department of physics, Graduate School, Inje University,
   Kimhae 621-749, Korea}
\end{center}
\vspace{5mm}
\centerline{{\bf{Abstract}}}
\vspace{5mm}

We obtain a Bekenstein entropy bound for the charged objects in arbitrary
dimensions ($D\ge 4$) using the D-bound recently proposed by Bousso.
With the help of thermodynamics of CFTs corresponding to AdS
Reissner-Norstr\"om (RN) black holes, we discuss the relation between
the Bekenstein and Bekenstein-Verlinde bounds. In particular we propose
a Bekenstein-Verlinde-like bound for the charged systems. In the
Einstein-Maxwell theory with a negative cosmological constant, we discuss
the brane cosmology with positive tension using the Binetruy-Deffayet-Langlois
approach. The resulting Friedman-Robertson-Walker equation can be identified
with the one obtained by the moving domain wall approach in the AdS RN black
hole background. Finally we also address the holographic property of the
brane universe.

\end{titlepage}

\newpage
\renewcommand{\thefootnote}{\arabic{footnote}}
\setcounter{footnote}{0}
\setcounter{page}{2}

\sect{Introduction}

In a recent paper \cite{Verl}, Verlinde made two interesting observations.
One is that according to the  AdS/CFT correspondence, the entropy of a
conformal field theory (CFT) in any dimension can be expressed in terms of
a generalized form of the  Cardy formula~\cite{Card}. Consider a certain CFT
residing in an $(n+1)$-dimensional spacetime with the metric
\begin{equation}
\label{1eq1}
ds^2 =-dt^2 +R^2d\Omega_n^2,
\end{equation}
where $d\Omega^2_n$ denotes the line element of a unit $n$-dimensional sphere.
It is proposed that the entropy of the CFT can be related to its total
energy $E$ and Casimir energy $E_c$ as
\begin{equation}
\label{1eq2}
S=\frac{2\pi R}{\sqrt{ab}}\sqrt{E_c(2E-E_c)}.
\end{equation}
Here $a$ and $b$ are two positive parameters. For strongly coupled CFTs with
the AdS duals, which implies that the CFTs are in the regime of supergravity
duals, $ab$ is fixed to be $n^2$ exactly. Thus one obtains the Cardy-Verlinde
formula
\begin{equation}
\label{1eq3}
S=\frac{2\pi R}{n}\sqrt{E_c(2E-E_c)}.
\end{equation}
Indeed, this formula holds for various kinds of AdS spacetimes:
AdS Schwarzschild black holes~\cite{Verl}; AdS Kerr black holes~\cite{Klem};
charged AdS black holes~\cite{Cai1}; Taub-Bolt AdS spacetimes~\cite{Birm1},
whose thermodynamics corresponds to that of different CFTs~\cite{Witten2}.

The other comes from the comparison of the Cardy formula with the
Friedman-Robertson-Walker (FRW) equation. For an $(n+1)$-dimensional
closed universe, the FRW equations are
\begin{eqnarray}
\label{1eq4}
&& H^2 =\frac{16 \pi G_n}{n(n-1)}\frac{E}{V} -\frac{1}{R^2}, \\
&& \dot{H}=-\frac{8 \pi G_n}{n-1}\left (\frac{E}{V} +p\right) +\frac{1}{R^2},
\end{eqnarray}
where $H=\dot{R}/R$ is the Hubble parameter, the dot stands for the
differentiation with respect to the proper time, $E$ is the total
energy of matter filling in the universe, $p$ denotes the pressure and
$V= R^n \mbox{Vol}(S^n)$ is the volume of the universe. In addition, $G_n$
is the $(n+1)$-dimensional gravitational constant. Verlinde pointed out that
the FRW equation~(\ref{1eq4}) can be related to three cosmological
entropy bounds:\footnote{In Ref.~\cite{Verl} the first bound is called
the Bekenstein bound. In fact, this bound is  slightly  different from
the original Bekenstein entropy bound \cite{Beke}. So we call this the
Bekenstein-Verlinde bound in our paper. We have more to comment on this point.}
\begin{equation}
\label{1eq6}
\begin{array}{rl}
\mbox{Bekenstein-Verlinde bound}: &
S_{\rm BV}=\frac{2\pi}{n}ER, \\
\mbox{Bekenstein-Hawking bound}: &
S_{\rm BH}=(n-1)\frac{V}{4G_nR}, \\
\mbox{Hubble  bound}: &
S_{\rm H}=(n-1)\frac{HV}{4G_n}.
\end{array}
\end{equation}
The FRW equation~(\ref{1eq4}) can then be rewritten as
\begin{equation}
\label{1eq7}
S_{\rm H}=\sqrt{S_{\rm BH}(2S_{\rm BV}-S_{\rm BH})}.
\end{equation}
The Bekenstein-Verlinde bound is valid for the weakly self-gravitating
universe ($HR<1$), while the Hubble bound holds for the strongly
self-gravitating universe ($HR>1$). It is clear from the FRW
equation~(\ref{1eq4}) that at the critical point of $HR=1$, three
entropy bounds coincide with each other.

Let us define $E_{\rm BH}$ corresponding to the Bekenstein-Hawking entropy
bound by the Bekenstein-Verlinde bound such that $S_{\rm BH}=(n-1) V/4G_nR
\equiv 2\pi E_{\rm BH} R/n$. The FRW equation~(\ref{1eq7}) then takes the form
\begin{equation}
\label{1eq8}
S_{\rm H}=\frac{2\pi R}{n}\sqrt{E_{\rm BH}(2E-E_{\rm BH})}.
\end{equation}
This relation has the same form as the Cardy-Verlinde formula~(\ref{1eq3}):
the entropy $S$ and the Casimir energy $E_c$ of CFTs are replaced by the
Hubble entropy bound $S_H$ and the black hole mass $E_{\rm BH}$ corresponding
to the Bekenstein-Hawking entropy bound.
This means that the FRW equation somehow knows the entropy of CFTs filling
in the universe~\cite{Verl}. This connection between the Cardy-Verlinde
formula and the FRW equation can be interpreted as a consequence of the
holographic principle~\cite{Verl}.

More recently, Savonije and Verlinde~\cite{Savo} have studied the brane
cosmology in the background of $(n+2)$-dimensional AdS Schwarzschild
spacetimes. It turns out that the equations governing the motion of
the brane are exactly the $(n+1)$-dimensional FRW equations with
radiation matter. This radiation matter can be viewed as the  CFT corresponding
to the black hole in the AdS/CFT correspondence. In addition, it is found 
that the FRW equation is exactly matched with the Cardy-Verlinde formula for
CFTs when the brane crosses the black hole horizon.

In this paper we extend this holographic connection to
the AdS Reissner-Nordstr\"om (RN) black hole background in arbitrary
dimensions. In the next section, we obtain a Bekenstein entropy bound for
charged objects in arbitrary dimensions ($D \ge 4$). This will be derived
by using the D-bound proposed by Bousso~\cite{Bous}. In Sec.~3, we briefly
review thermodynamic aspects of the AdS RN black holes.  We
discuss the relation between the Bekenstein and Bekenstein-Verlinde bounds.
In particular, we propose a Bekenstein-Verlinde-like bound for charged
systems.

In Sec.~4 we consider the cosmology of brane with positive tension in the
Einstein-Maxwell theory with a negative cosmological constant using the
Binetruy-Deffayet-Langlois (BDL)
approach~\cite{BDL1,BDL2}. We find that the resulting FRW equation can be
identified with that obtained by Biswas and Mukherji~\cite{BM} by using
the moving domain wall approach~\cite{Krau,Ida,Nun}. Finally we address the
holographic property of this brane cosmology. The conclusions are given
in Sec.~5.

Other related discussions to Verlinde's observations can be found in
Refs.~\cite{Kuta}-\cite{Youm}.


\sect{Bekenstein bound in arbitrary dimensions }

Bekenstein~\cite{Beke} is the first to consider the issue of maximal entropy
for a macroscopic system. He argued that for a closed system with total
energy $E$, which fits in a sphere with radius $R$ in three spatial
dimensions, there exists an upper bound on the entropy
\begin{equation}
\label{2eq1}
S \le S_{\rm B}= 2\pi R E.
\end{equation}
This is called the Bekenstein entropy bound.

The Bekenstein bound is believed to be valid for a system with the limited
self-gravity, which means that the gravitational self-energy is negligibly
small compared to its total energy. However, it is interesting to note that
the bound~(\ref{2eq1}) is saturated even for a four-dimensional Schwarzschild
black hole which is a strongly self-gravitating object. Furthermore, it has
been found in Ref.~\cite{Bous} that the form of the Bekenstein
bound~(\ref{2eq1}) is independent of the spatial dimensions. This was
obtained by considering the Geroch process in a $D (\ge 4)$-dimensional
Schwarzschild background and the generalized second law of black hole
thermodynamics. It implies that the Bekenstein bound in arbitrary dimensions
($D \ge 4)$ always remains in the same form~(\ref{2eq1}). It is easy to
check that a $D (\ge 4)$-dimensional Schwarzschild black hole satisfies the
bound (\ref{2eq1}). But the bound will no longer be saturated for $D>4$.

The bound~(\ref{2eq1}) has been extended recently to  the case of
 charged objects in four dimensions~\cite{Hod,Mayo,Linet1,Linet2}.
It is found that the Bekenstein bound is modified to
\begin{equation}
\label{2eq2}
S \le S_{\rm B}=\pi (2ER -Q^2),
\end{equation}
for a closed system with charge $Q$. This bound is saturated by
a four-dimensional RN black hole with mass $E$ and charge $Q$.

It is very interesting to investigate whether or not the form (\ref{2eq2})
of the Bekenstein bound for a charged object in arbitrary dimensions
($D\ge 4$) remains unchanged  as in the case for the neutral objects.
 Furthermore, we
need such a bound in order to discuss the holographic aspects of the brane
universe in the AdS RN black hole background. An important ingredient in
deriving the Bekenstein bound~(\ref{2eq2}) is the electrostatic
self-energy of a charged test object in a black hole
background~\cite{Hod,Mayo,Linet1,Linet2}. However, it is not easy to obtain
this quantity in higher dimensions. So here we use the D-bound proposed by
Bousso~\cite{Bous} to get a Bekenstein bound for a charged system
in arbitrary dimensions.

We start with the $(n+2)$-dimensional Einstein-Maxwell theory with a
cosmological constant $\Lambda_{\pm} = \pm n(n+1)/2l^2$:
\begin{equation}
\label{2eq3}
I = \frac{1}{16\pi \g}\int d^{n+2}x\sqrt{-g} \left ( {\cal R}
-F_{\mu\nu}F^{\mu\nu} - 2 \Lambda_{\pm}\right),
\end{equation}
where ${\cal R}$ is the curvature scalar, $F$ denotes the Maxwell field, and
$\g$ is the gravitational constant in ($n+2)$ dimensions.
Varying the action (\ref{2eq3}) yields the equations of motion
\begin{eqnarray}
\label{2eq4}
&& G_{\mu\nu} \equiv {\cal R}_{\mu\nu} -\frac{1}{2}g_{\mu\nu}{\cal R}
   = T_{\mu\nu}, \ \ \
T_{\mu\nu}=  2F_{\mu\lambda}F_{\nu}^{\ \lambda}-\frac{1}{2}
 g_{\mu\nu} F^2 - \Lambda_{\pm}g_{\mu\nu}, \\
\label{2eq5}
&& \partial_{\mu}(\sqrt{-g}F^{\mu\nu})=0.
\end{eqnarray}
These equations have a spherically symmetric solution~\cite{CS,Cham}
\begin{eqnarray}
\label{2eq6}
 && ds^2 = -f(r)dt^2 +f(r)^{-1}dr^2 +r^2 d\Omega_n^2, \nonumber \\
\label{2eq7}
 && F_{rt}=\frac{n\omega_n}{4}\frac{Q}{r^n}, \ \ \ \
 \omega_n=\frac{16\pi \g}{n \mbox{Vol}(S^n)},
\end{eqnarray}
where $\mbox{Vol}(S^n)$ is the volume of a unit $n$-sphere and the function
$f$ is given by
\begin{equation}
\label{2eq8}
f_{\pm}(r) = 1 -\frac{\omega_n M}{r^{n-1}} +\frac{n \omega^2_n Q^2}{8(n-1)
   r^{2(n-1)}} -\frac{2\Lambda_{\pm}r^2}{n(n+1)}.
\end{equation}
This solution is asymptotically de Sitter (dS) or anti-de Sitter (AdS)
depending on the cosmological constant $\Lambda_+$ or $\Lambda_-$.
In this section we consider the de Sitter case.

We note that if $M=Q=0$, the solution reduces to the dS space which has
a cosmological horizon at $r=r_0\equiv \sqrt{l^2}$. The cosmological
horizon behaves in many aspects like the black hole horizon~\cite{Gibb}.
In particular, it has the thermodynamic entropy
\begin{equation}
\label{2eq9}
S_0=\frac{r_0^n}{4\g}\mbox{Vol}(S^n).
\end{equation}
In a more general case with nonvanishing $M$ and $Q$, the solution
 describes the geometry of
a certain object with mass $M$ and electric charge $Q$ embedded in dS
space\footnote{In asymptotic dS space, the energy of the system is not
well-defined due to the absence of a suitable spacelike infinity. Here
the constant $M$ is viewed as the mass of the object in the sense of
Ref.~\cite{Abbo}.}. The cosmological horizon will shrink due to the nonzero
$M$ and $Q$. This leads to the $N$-bound of Bousso~\cite{Bous}. This bound
claims that in the asymptotically dS spacetime, the maximally observable 
degrees of freedom are bounded by the entropy~(\ref{2eq9}) of the exact dS 
space.
Furthermore, applying the Geroch process to the cosmological horizon leads
to the D-bound in dS space~\cite{Bous}. This tells us that the entropy of
objects in dS space is bounded by the difference of the entropies in
the exact dS space and in the asymptotically dS space
\begin{equation}
\label{2eq10}
S_m \le S_0 -S_c,
\end{equation}
where $S_c$ is the cosmological horizon entropy when matter is present.

Let us apply this D-bound to the dS RN spacetime~(\ref{2eq6}). Here the
cosmological horizon $r_c$ is given by the maximal root of the equation: 
\begin{equation}
\label{2eq11}
1-\frac{\omega_n M}{r_c^{n-1}} +\frac{n\omega_n^2 Q^2}{8(n-1)r_c^{2(n-1)}}
   -\frac{r_c^2}{r_0^2}=0.
\end{equation}
This leads to
\begin{equation}
\label{2eq12}
\frac{r_0^n}{r_c^n} =\left (1 -\frac{\omega_n M}{r_c^{n-1}}
  +\frac{n\omega_n^2 Q^2}{8 (n-1)r_c^{2(n-1)}} \right)^{-n/2}.
\end{equation}
Consider the large cosmological horizon limit of $M/r_c^{n-1} \ll 1$
and $Q^2/r_c^{2(n-1)} \ll 1$,  we have
\begin{equation}
\frac{r_0^n}{r_c^n} \approx 1 +\frac{n\omega_n M}{2r_c^{n-1}}
   -\frac{n^2\omega_n^2 Q^2}{16(n-1)r_c^{2(n-1)}},
\end{equation}
in the leading order of $M/r_c^{n-1}$ and $Q^2/r_c^{2(n-1)}$. Substituting
the above  to the D-bound (\ref{2eq10}) gives
\begin{eqnarray}
\label{2eq13}
S_m &\le & \frac{\mbox{Vol}(S^n)}{4\g} (r_0^n-r_c^n) \nonumber \\
 &\le& 2\pi r_c \left (M -\frac{2\pi {\g} Q^2}{(n-1)\mbox{Vol}(S^n) r_c^{n-1}}
  \right ).
\end{eqnarray}
One finds that the entropy reaches its maximum  when the matter extends to the
cosmological horizon since the distribution range of matter is bounded by the
cosmological horizon. Replacing $r_c$ by $R$ and $M$ by the proper energy $E$,
we get an entropy bound of the charged object in arbitrary dimensions
($D=n+2 \ge 4 $):
\begin{equation}
\label{2eq14}
S \le S_{\rm B}= 2\pi R \left (E -\frac{2\pi {\g} Q^2}{(n-1) V(S^n)R^{n-1}}
     \right).
\end{equation}
To check this result (\ref{2eq14}), we note that
when $Q^2=0$, this bound reproduces precisely the Bekenstein bound~(\ref{2eq1})
for the neutral object in arbitrary dimensions. For $n=2$, the entropy
bound~(\ref{2eq14}) reduces to the four-dimensional one~(\ref{2eq2}).\footnote{
In the bound~(\ref{2eq2}), the gravitational constant $\g$ is absent, but it
appears in (\ref{2eq14}). This is due to the different units for electric
charge used in~\cite{Hod,Mayo,Linet1,Linet2} and in this paper.}
Furthermore, comparing (\ref{2eq14}) and (\ref{2eq2}), we note that unlike the
 neutral case, the Bekenstein bound
for charged objects in higher dimensions ($D>4$) changes its form from that
in four dimensions. In addition, one can see that the entropy 
bound (\ref{2eq14}) will no longer be saturated by a higher ($D>4$) 
dimensional RN black hole. This is the same as the case of neutral objects.

We can rewrite the bound (\ref{2eq14}) in a more holographic form:
\begin{equation}
\label{2eq15}
S_{\rm B} = 2\pi R\left(E -\frac{2\pi {\g} R Q^2}{(n-1) {\cal A}} \right),
\end{equation}
where ${\cal A}=R^n\mbox{Vol}(s^n)$ is the surface area of the charged 
object. The second term in the r.h.s. of Eq.~(\ref{2eq15}) can be understood 
as the energy $E_q$ of the electromagnetic field outside the charged object.
Hence one can also write
\begin{equation}
\label{2eq16}
S_{\rm B}= 2\pi R(E-E_q), \ \ \  E_q =\frac{1}{2}\phi Q,
\end{equation}
where $\phi = \frac{n \omega_n}{4(n-1)}\frac{Q}{R^{n-1}}$ is the electrostatic
potential at the surface of the charged object. In deriving Eq.~(\ref{2eq16}),
we have assumed that the potential vanishes at the spatial infinity.

\sect{AdS Reissner-Nordstr\"om black holes and
     Bekenstein-Verlinde bound}

In this section we consider the AdS case of $\Lambda_-$ in the
solution~(\ref{2eq6}). The cosmological horizon is absent in this case and
the solution (\ref{2eq6}) describes the AdS RN black hole in arbitrary
dimensions. The black hole horizon $r_+$ is determined by the maximal root
of the equation $f_-(r_+)=0$. The integration constants $M$ and $Q$ can be
interpreted as the mass and electric charge of the black hole.

In the spirit of the AdS/CFT correspondence, the thermodynamics of AdS RN
black holes corresponds to that for the boundary CFT with an R-charge
(or R-potential). In Ref.~\cite{Cai1}, it was shown that indeed the entropy of
the corresponding CFTs can be expressed in terms of the Cardy-Verlinde
form~(\ref{1eq3}). In this section we further discuss aspects of the
thermodynamic properties and suggest a Bekenstein-Verlinde-like bound for
the charged systems.

We rescale the boundary metric so that it has the form~(\ref{1eq1}).
Thermodynamic quantities of the corresponding CFT are given by
\begin{eqnarray}
\label{3eq1}
&& E = \frac{l r_+^{n-1}}{R\omega_n}\left ( 1+\frac{r_+^2}{l^2}
       +\frac{n\omega_n^2 Q^2}{8(n-1)r_+^{2(n-1)}}\right), \nonumber\\
&& T =\frac{l}{4\pi Rr_+}\left((n-1) +\frac{(n+1)r_+^2}{l^2}
    -\frac{n\omega_n^2 Q^2}{8r_+^{2(n-1)}} \right), \nonumber \\
&& \Phi =\frac{nl\omega_n Q}{4(n-1)R r_+^{n-1}}, \nonumber \\
&& S = \frac{r_+^n}{4\g}\mbox{Vol}(S^n), \nonumber \\
&& G = \frac{lr_+^{n-1}}{nR\omega_n}\left(1-\frac{r_+^2}{l^2}
     -\frac{n\omega_n^2}{8(n-1)}\frac{Q^2}{r_+^{2(n-1)}}\right),
\end{eqnarray}
where $r_+$ is the horizon of the AdS RN black hole, and $E$, $T$, $\Phi$,
$S$ and $G$ represent the energy, temperature, chemical potential conjugate
to the charge $Q$, entropy and Gibbs free energy of the CFT, respectively.
In analogy to the Cardy-Verlinde formula~(\ref{1eq3}), we can rewrite the
entropy in Eq.~(\ref{3eq1}) as
\begin{equation}
\label{3eq2}
S =\frac{2\pi R}{n}\sqrt{E_c(2(E-E_q)-E_c}),
\end{equation}
where
\begin{equation}
\label{3eq3}
E_c = \frac{2lr_+^{n-1}}{\omega_n R}, \ \ \  E_q=\frac{1}{2}\Phi Q =
   \frac{l}{R}\frac{n\omega_n}{8(n-1)}\frac{Q^2}{r_+^{n-1}}.
\end{equation}
We note that $E_q$ is the  bulk energy of electromagnetic field $E_B =
-\frac{1}{16\pi \g} \int^{\infty}_{r_+}d^{n+2}x\sqrt{-g}F^2$ multiplied
by the scale factor $(l/R)$.  For the fixed $E$, $R$ and $E_q$, we find
from (\ref{3eq2}) that  the entropy  reaches its maximal value
\begin{eqnarray}
\label{3eq5}
S_{\rm max} &=& \frac{2\pi R}{n}(E-E_q) \nonumber \\
            &=& \frac{2\pi}{n}\left (ER -\frac{nl\omega_n}{8(n-1)}
              \frac{Q^2}{r_+^{n-1}}\right),
\end{eqnarray}
at $E_c =E-E_q$. Note that the above expression is quite similar to the
Bekenstein bound~(\ref{2eq16}) for the charged objects.

Now let us discuss the difference and relation between the Bekenstein-Verlinde
bound $S_{\rm BV}$ in (\ref{1eq6}) and the Bekenstein bound (\ref{2eq1}).
In the AdS/CFT correspondence, the CFT resides in the UV boundary  of the
AdS spacetime and the gravity decouples on the boundary. Recall that the
Cardy-Verlinde formula~(\ref{1eq3}) is supposed to give the entropy of the
($n+1$)-dimensional CFT residing in the spacetime~(\ref{1eq1}), UV boundary
of the AdS space. It holds at
least in the regime of supergravity duals. Note further that the
Cardy-Verlinde formula~(\ref{1eq3}) gives the maximal entropy
($S=2\pi E R/n$), when the Casimir energy equals the total energy ($E_c=E$).
This is just the Bekenstein-Verlinde bound $S_{\rm BV}$.
Therefore it is reasonable to regard the Bekenstein-Verlinde bound
$S_{\rm BV}$ as the maximal entropy bound of CFTs in (\ref{1eq1}), rather
than a certain entropy bound of a system with gravity.  On the other
hand, the Bekenstein bound~(\ref{2eq1}) for neutral objects  holds for
a closed system in an asymptotically spacetime.  Therefore, the Beksentein
bound and the Bekenstein-Verlinde bound are applicable in different spacetimes. 
Furthermore, the Bekenstein bound for neutral objects  remains in the
same form~(\ref{2eq1}) in any dimensions~\cite{Bous}, while the form of the
Bekenstein-Verlinde bound depends on the spacetime dimension ($n$).

Let us consider the brane world scenario~\cite{Gubser} in the
generalized AdS/CFT correspondence. The gravity does not
decouple on the brane because the brane is not on the UV boundary of AdS
space but it is located in the bulk of the AdS space. Like the Bekenstein
bound (\ref{2eq1}), suppose the Bekenstein-Verlinde bound in (\ref{1eq6}) 
is also valid for CFTs with limited self-gravity. Thus we can regard
the Bekenstein-Verlinde bound $S_{\rm BV}$ in (\ref{1eq6}) as a counterpart 
(on the brane) of the Bekenstein bound~(\ref{2eq1}), because the reduced metric
on the brane is of the form (\ref{1eq1}). 
That is, the Bekenstein bound is valid in the bulk, while the 
Bekenstein-Verlinde bound holds on the
brane.  Of course, for both cases, the gravity is assumed to be weak.

The FRW cosmology renders a piece of evidence for supporting this.
We know that the Bekenstein bound (\ref{2eq1}) cannot be naively used for 
a closed universe because of the lack of a suitable boundary~\cite{Verl}. 
On the other
hand, the Bekenstein-Verlinde bound holds in the spacetime~(\ref{1eq1}).
This has a scale $R$, which can be naturally identified with the cosmic scale
$R$ in the FRW cosmology. Furthermore, the Bekenstein-Verlinde bound
$S_{\rm BV}$ appears in the FRW equation. Thus, we can also view
the Bekenstein-Verlinde bound as a counterpart of the Bekenstein bound
in the context of brane cosmology.

With these considerations, we now suggest a Bekenstein-Verlinde-like bound
for a charged system.
According to the AdS/CFT correspondence, the boundary spacetime in which
the boundary CFT resides can be determined from the bulk metric, up to a
conformal factor. The conformal factor enables us to rescale the boundary
metric as we wish. In order to obtain a suitable bound, we
rescale the boundary metric so that the radius $R$ in (\ref{1eq1}) becomes
the horizon radius $r_+$ of the black hole, as in~\cite{Verl}. 
The maximal entropy (\ref{3eq5})  is then given by
\begin{equation}
\label{3eq6}
S_{\rm max}=\frac{2\pi R}{n}\left (E -\frac{nl\omega_n}{8(n-1)}\frac{Q^2}
   {R^n}\right).
\end{equation}
Consider further the similarity between the Bekenstein bound (\ref{2eq1})
for neutral objects  and the Bekenstein-Verlinde bound $S_{\rm BV}$ in 
(\ref{1eq6}), and also the similarity between the Bekenstein 
bound~(\ref{2eq15}) for charged objects and the maximal 
entropy~(\ref{3eq5}) of the CFT with R-charge.
We propose the maximal entropy (\ref{3eq6}) as the Bekenstein-Verlinde-like
bound $S_{\rm BV}$ for the
CFT with the R-charge $Q$ in the spacetime~(\ref{1eq1}). We can rewrite
the above as
\begin{equation}
\label{3eq7}
S_{\rm BV} =\frac{2\pi R}{n}\left (E -\frac{2\pi {\g}l}{(n-1)}\frac{Q^2}{V}
        \right),
\end{equation}
where $V= R^n \mbox{Vol}(S^n)$. 
One may wonder why the bulk parameter $l$ appears in the above bound.
This can be understood by recalling that in the AdS/CFT correspondence,
the cosmological constant is related to the 't Hooft coupling constant
in the CFT. Furthermore, we will see that the bound (\ref{3eq7}) plays the
same role in the brane cosmology in the AdS RN black hole background as
$S_{\rm BV}$ in (\ref{1eq6}) in the FRW equation (\ref{1eq7}).


\sect{Brane cosmology in the charged background}

Recently the cosmology of the Randall-Sundrum (RS) scenario~\cite{RS}
for a positive tension brane in a five-dimensional universe (with localized
gravity) has been studied extensively. In most of works  a negative
cosmological constant is introduced as the bulk matter without else.
In this section we consider a higher dimensional cosmology of RS scenario
with a Maxwell field as well as the negative cosmological constant as
the bulk matter.  That is, we consider a brane universe in the bulk with 
the action~(\ref{2eq3}).

Following BDL~\cite{BDL1,BDL2}, we assume the bulk metric is of the
form
\begin{equation}
\label{4eq1}
ds^2 = -c^2(t,y) dt^2 +a^2(t,y)\gamma_{ij}dx^idx^j +b^2(t,y)dy^2,
\end{equation}
where $\gamma_{ij}$ is the metric of an $n$-dimensional space with constant
curvature $n(n-1)k$. One may take $k=1$, $0$ and $ -1$.  In the orthogonal 
basis, we work out the
Einstein tensor
\begin{eqnarray}
\label{4eq2}
G_{\hat t \hat t} &=& n \left [\frac{\dot a}{ac^2}\left(\frac{n-1}{2}
    \frac{\dot a}{a}+\frac{\dot b}{b}\right)
   -\frac{1}{b^2}\left (\frac{a''}{a}+\frac{a'}{a}\left( \frac{n-1}{2}
    \frac{a'}{a}-\frac{b'}{b}\right) \right)
   +\frac{n-1}{2}\frac{k}{a^2}\right ], \nonumber \\
G_{\hat y \hat y} &=&n \left [ \frac{a'}{ab^2}\left(
   \frac{n-1}{2}\frac{a'}{a} +\frac{c'}{c} \right)
    -\frac{1}{c^2}\left( \frac{\ddot a}{a}
    +\frac{\dot a}{a}\left ( \frac{n-1}{2}\frac{\dot a}{a}-\frac{\dot c}{c}
    \right)\right) -\frac{n-1}{2}\frac{k}{a^2}\right], \nonumber \\
G_{\hat t \hat y} &=& n \left (\frac{\dot a c'}{abc^2}
                  + \frac{a'\dot b}{ab^2 c} -\frac{\dot a'}{abc} \right),
      \nonumber \\
G_{\hat i \hat j} &=& \frac{\delta_{ij}}{b^2} \left[ (n-1)\frac{a''}{a}
          +\frac{c''}{c}+\frac{n-1}{2}\frac{a'}{a}
            \left((n-2)\frac{a'}{a} +2\frac{c'}{c}
           \right) -\frac{b'}{b}\left ((n-1)\frac{a'}{a}
        +\frac{c'}{c}\right) \right] \nonumber \\
     & & + \frac{\delta_{ij}}{c^2}\left [-(n-1)\frac{\ddot a}{a}
     -\frac{\ddot b}{b} +\frac{\dot b}{b}\left (\frac{\dot c}{c}- (n-1)
      \frac{\dot a}{a}\right)
     +\frac{n-1}{2} \frac{\dot a}{a} \left (2 \frac{\dot c}{c} -(n-2)
         \frac{\dot a}{a}\right) \right] \nonumber \\
    & & -\frac{(n-1)(n-2)}{2}\frac{k}{a^2}\delta_{ij},
\end{eqnarray}
where the dot (prime) stands for the differentiation with respect to $t$ ($y$).

Now suppose that an $n$-dimensional brane is located at $y=0$. On the two
sides of the brane, the stress-energy tensors are given by~(\ref{2eq4}),
but they are not necessarily identified. For example, the cosmological
constant may be different on two sides~\cite{Tye}. The stress-energy
tensor on the brane is assumed to be of the form
\begin{equation}
\label{4eq3}
\tau_{\mu}^{\ \nu} = \frac{\delta(y)}{b}{\rm diag}(-\rho,p,\cdots,p,0),
\end{equation}
which implies that the brane is homogeneous and isotropic.

Let us denote the gap of a given function $f$ at $y=0$ by
$[f]=f(0_+) -f(0_-)$ and its average by $\{f\}=(f(0_+) +f(0_-))/2$.
The functions $a$, $b$, $c$ in the metric~(\ref{4eq1}) are continuous
at $y=0$, but their derivatives are discontinuous. So the second derivatives
are of the form~\cite{Tye}
\begin{equation}
\label{4eq4}
f''=f''|_{(y\ne 0)} +[f']\delta(y).
\end{equation}
It is then straightforward to write down the gaps in the ($tt$), ($yy$)
and ($ij$) components of the Einstein equation~(\ref{2eq4}), respectively:
\begin{eqnarray}
\label{4eq5}
&& \frac{n}{b^2_0}\left (- (n-1)\frac{[a']\{a'\}}{a^2_0}
   +\frac{[a']\{b'\}}{a_0b_0} +\frac{\{a'\}[b']}{a_0b_0}\right)
    = T_{\hat t \hat t}(0_+) -T_{\hat t \hat t}(0_-),
     \nonumber \\
&& \frac{n}{b_0^2}\left ( (n-1)\frac{[a']\{a'\}}{a_0^2}
    +\frac{[a']\{c'\}}{a_0c_0} +\frac{\{a'\}[c']}{a_0c_0}
     \right) = T_{\hat y \hat y}(0_+) -T_{\hat y \hat y}(0_-),
   \nonumber \\
&& \frac{(n-1)}{b_0^2}\delta_{ij} \left( (n-2)\frac{[a']\{a'\}}{a_0^2}
   +\frac{[a']\{c'\}}{a_0c_0} +\frac{\{a'\}[c']}{a_0c_0}
   -\frac{[a']\{b'\}}{a_0b_0}-\frac{\{a'\}[b']}{a_0b_0} \right. \nonumber \\
&& ~~~~~~~~~~~~\left. -\frac{1}{n-1}\frac{[b']\{c'\}}{b_0c_0}
     -\frac{1}{n-1}\frac{\{b'\}[c']}{b_0c_0} \right )=T_{\hat i \hat j}(0_+)
   -T_{\hat i \hat j}(0_-).
\end{eqnarray}
where the quantities with subscript $0$ denote those at $y=0$. The
$\delta$-function parts in the ($tt$) and ($ij$) components of the Einstein
equation give
\begin{equation}
\label{4eq6}
\frac{[a']}{a_0b_0}=-\frac{8\pi \g}{n}\rho, \ \  \  \frac{[c']}{b_0c_0}
   =8 \pi{\g}\left (p +\frac{n-1}{n}\rho\right).
\end{equation}
On the other hand, the average part of the ($yy$)-component is
\begin{eqnarray}
\label{4eq7}
&& \frac{1}{c_0^2}\left (\frac{\ddot a_0}{a_0} +\frac{\dot a_0}{a_0}
   \left(\frac{n-1}{2} \frac{\dot a_0}{a_0} -\frac{\dot c_0}{c_0}\right)
   \right)
=-\frac{1}{2n}\left (T_{\hat y\hat y}(0_+) +
        T_{\hat y \hat y}(0_-)\right)   \nonumber \\
 && ~~~~~ +\frac{n-1}{2}\left (-\frac{k}{a^2_0}
    +\frac{1}{4}\left (\frac{[a']}{a_0b_0}\right)^2
   +\left(\frac{\{a'\}}{a_0b_0}\right)^2 \right)
   +\frac{1}{4} \frac{[a'][c']}{a_0b_0^2 c_0}
  +\frac{\{a'\}\{c'\}}{a_0b_0^2 c_0}.
\end{eqnarray}
We note that the Maxwell equation~(\ref{2eq5}) in the metric~(\ref{4eq1})
has the solution
\begin{equation}
\label{4eq8}
F_{yt} =\frac{{\cal Q}bc}{a^n},
\end{equation}
where ${\cal Q}$ is an integration constant. Thus the bulk stress-energy
tensor including the Maxwell field and cosmological constant $\Lambda_-$ is
\begin{equation}
\label{4eq9}
T_{\hat \mu \hat \nu} ={\rm diag}\left (-\frac{n(n+1)}{2l^2}
  +\frac{{\cal Q}^2}{a^{2n}}, \frac{n(n+1)}{2l^2} +\frac{{\cal Q}^2}{a^{2n}},
   \cdots, \frac{n(n+1)}{2l^2} +\frac{{\cal Q}^2}{a^{2n}},
   \frac{n(n+1)}{2l^2}-\frac{{\cal Q}^2}{a^{2n}} \right).
\end{equation}

Now we consider a simple case in which the bulk is $Z_2$-symmetric and the
bulk stress-energy tensors are identical on two sides of the brane. One
then has $\{f'\}=0$. We define the cosmic time $\tau$ as
$d\tau =c(t,0)dt$ and
the Hubble parameter $H$ on the brane at $y=0$ as $H =\dot a_0/a_0 =\dot R/R$.
Following~\cite{Tye}, the l.h.s. of Eq.~(\ref{4eq7}) can be rewritten as
\begin{equation}
\frac{1}{c_0^2}\left (\frac{\ddot a_0}{a_0} +\frac{\dot a_0}{a_0}
   \left(\frac{n-1}{2} \frac{\dot a_0}{a_0} -\frac{\dot c_0}{c_0}\right)
   \right) = \frac{1}{2R^n}\frac{d}{dR}H^2R^{n+1},
\end{equation}
while the r.h.s. can be calculated with the help of Eqs.~(\ref{4eq6}) and
(\ref{4eq9}). Then (\ref{4eq7}) leads to
\begin{equation}
\label{4eq11}
\frac{1}{2R^n}\frac{d}{dR}H^2R^{n+1}= -\frac{(8\pi \g)^2 }{8n^2}\rho\left(
     2n p +(n-1)\rho \right) -\frac{n+1}{2l^2}
    -\frac{n-1}{2}\frac{k}{R^2} +\frac{{\cal Q}^2}{n R^{2n}}.
\end{equation}
Without the localized matter on the brane, we have $p=-\rho =-n/4\pi l\g$.
 Here the tension of the brane is assumed to be fine-tuning.
Thus Eq.~(\ref{4eq11}) reduces to
\begin{equation}
\label{4eq12}
\frac{d}{dR}H^2 R^{n+1}=-(n-1)k R^{n-2} +\frac{2{\cal Q}^2}{nR^n}.
\end{equation}
Integrating this equation yields
\begin{equation}
\label{4eq13}
H^2 = \frac{{\cal C}}{R^{n+1}} -\frac{k}{R^2} -\frac{2}{n(n-1)}
\frac{{\cal Q}^2}{R^{2n}},
\end{equation}
where ${\cal C}$ is an integration constant, which already appears
in~\cite{BDL2}. This is our  equation derived from the BDL approach
and governs the evolution of
the brane universe in the bulk with a Maxwell field.

Now let us  compare our equation (\ref{4eq13}) with the equation of motion
for the moving domain wall (brane) in the AdS RN black hole
solutions~(\ref{2eq6}). The latter is~\cite{BM}
\begin{equation}
\label{4eq14}
H^2 =\frac{\omega_n M}{R^{n+1}}-\frac{1}{R^2}
 -\frac{n\omega^2_n Q^2}{8(n-1)R^{2n}}.
\end{equation}
Taking $k=1$ and 
\begin{equation}
\label{4eq15}
{\cal C}= \omega_n M, \ \ \ \ {\cal Q}^2 = \frac{n^2 \omega_n ^2 Q^2}{16},
\end{equation}
one can see immediately that our equation (\ref{4eq13}) exactly coincides 
with Eq.~(\ref{4eq14}). For Eq.~(\ref{4eq14})
of the moving brane, the bulk is the AdS RN black hole geometry with
the mass $M$ and electric charge $Q$. These two parameters encode the
information of the bulk and describe the CFT on the brane.
On the other hand, in deriving Eq.~(\ref{4eq13}), we do not have to know
exactly what is the bulk geometry but we only have to know that there exists
a Maxwell field in the bulk. However, the two integration constants
${\cal C}$ and ${\cal Q}$ certainly encode the information of the bulk.
As a result, just as in the moving brane approach, in which the brane moves
in the bulk and acts as the boundary of the bulk, the FRW equation of the
brane in the BDL approach with fixed brane in the bulk, also encodes the
information of the bulk, showing the same holographic property.

There is a relation between the gravitational constants in the bulk
and on the brane~\cite{Gubser,Savo}:
\begin{equation}
{\g} =\frac{G_n l}{n-1}.
\end{equation}
Relating the mass $M$ of black hole with the energy on the brane, one can
rewrite (\ref{4eq14}) as
\begin{equation}
\label{4eq17}
H^2 = -\frac{1}{R^2} +\frac{16\pi G_n}{n(n-1)}\frac{E}{V}
    -\frac{32\pi^2 {\g}G_n l}{n(n-1)^2}\frac{Q^2}{V^2},
\end{equation}
where $E=lM/R$ and $V=R^n \mbox{Vol}(S^n)$. With the Bekenstein-Hawking
bound $S_{\rm BH}$, Hubble bound $S_{\rm H}$ in (\ref{1eq6}) and the
Bekenstein-Verlinde-like bound~(\ref{3eq7}) we proposed in the previous
section, we find that the FRW
equation~(\ref{4eq17}) can be expressed as
\begin{equation}
\label{4eq18}
S_{\rm H}=\sqrt{S_{\rm BH} (2S_{\rm BV}-S_{\rm BH})},
\end{equation}
which is the same as~(\ref{1eq7}), although the $S_{\rm BV}$ in (\ref{4eq18})
is  different from the one in (\ref{1eq7}).
 With the Bekenstein-Hawking energy $E_{\rm BH}$ defined in (\ref{1eq8},
 Eq.~(\ref{4eq18}) can be further cast into
\begin{equation}
\label{4eq19}
S_{\rm H}= \frac{2\pi R}{n}\sqrt{E_{\rm BH}\left (2\left (E-\frac{2\pi l {\g}}
    {n-1}\frac{Q^2}{V}\right) -E_{\rm BH}\right)}.
\end{equation}
This is of the same form as the Cardy-Verlinde formula~(\ref{3eq2}) for
the CFT with an R-charge corresponding to the AdS RN black holes. In 
particular, when the brane crosses the horizon of AdS RN black holes, 
{\it i.e.}, $R=r_+$, Eq.~(\ref{4eq19}) matches with the Cardy-Verlinde 
formula~(\ref{3eq2}) exactly.

\sect{Conclusions}
Using the D-bound of Bousso we have obtained a Bekenstein entropy
bound (\ref{2eq14}) for charged systems in arbitrary dimensions ($D\ge 4)$. 
When the charge vanishes, the bound reduces to the usual Bekenstein bound for 
neutral objects, while if one puts $D=4$, it precisely reproduces the known
Bekenstein entropy bound for charged objects in four dimensions. The 
Bekenstein entropy bound (\ref{2eq14}) is saturated by a four dimensional
RN black hole.  But as the case of neutral objects, the Bekenstein bound
will no longer be saturated by higher ($D>4)$ dimensional RN black holes.

We have also discussed the difference and relation between the Bekenstein
and Bekenstein-Verlinde bounds, and argued that the Bekenstein-Verlinde
bound could be regarded as the counterpart of the Bekenstein
bound in the context of cosmology. With the thermodynamics of AdS RN black 
holes, a Bekenstein-Verlinde-like bound for charged system has been suggested.
 In addition, we have studied the brane cosmology of RS scenario in the
Einstein-Maxwell theory with a negative cosmological constant in the BDL 
approach. The resulting FRW equation
can be identified with the one which governs the motion of a domain wall
in the AdS RN black hole background.  With the Bekenstein-Verlinde-like
bound and others, our FRW equation can be cast into the form of
Cardy-Verlinde formula, which describes the entropy of CFT with an R-charge 
filling  the
brane universe. Our results further indicate that the brane cosmologies
resulting from the BDL approach and from the moving brane approach can be
identified with each other, and show the same holographic properties in 
the case of Einstein-Maxwell theory.
This paper also supports and extends the result of \cite{Muko}. There the
authors discussed the relation between the BDL approach and moving
domain wall approach in the case where the bulk has only
a negative cosmological constant.

\section*{Acknowledgments}

The work of RGC and NO was supported in part by the Japan Society for the
Promotion of Science and in part by Grants-in-Aid for Scientific Research
Nos. 99020, 12640270 and in part by Grant-in-Aid on the Priority Area:
Supersymmetry and Unified Theory of Elementary particles. YSM was supported
in part by the Brain Korea 21 Program, Ministry of Education, Project No.
D-0025 and KOSEF, Project No. 2000-1-11200-001-3.


\end{document}